\documentclass{endm}
\usepackage{endmmacro}
\usepackage{mathrsfs,hyperref,xcolor}
\usepackage{amsmath,amsfonts,amssymb}

\def \ie {i.e.~}

\begin{document}

\begin{verbatim}\end{verbatim}\vspace{2.5cm}

\begin{frontmatter}

\title{An exact DSatur-based algorithm for the Equitable Coloring Problem}

\author[a]{Isabel M\'endez-D\'iaz\thanksref{GRANT}}
\author[b]{Graciela Nasini\thanksref{GRANT}}
\author[b]{Daniel Sever\'in\thanksref{GRANT}}

\address[a]{ FCEyN, Universidad de Buenos Aires,
	Argentina, \texttt{imendez@dc.uba.ar} }

\address[b]{ FCEIA, Universidad Nacional de Rosario y CONICET
	Argentina, \\ \texttt{\{nasini, daniel\}@fceia.unr.edu.ar} }

\thanks[GRANT]{Partially supported by grants 
PIP-CONICET 241, PICT 2011-0817 and UBACYT 20020100100666.}

\begin{abstract}
This paper describes an exact algorithm for the Equitable Coloring Problem, based on the well known \textsc{DSatur} algorithm for the classic Coloring Problem with new pruning rules specifically derived from the equity constraint.
Computational experiences 
show that our algorithm is competitive with those known in literature.
\end{abstract}


\end{frontmatter}


\vspace{-18pt}
\section{Introduction}

\vspace{-5pt}
The Graph Coloring Problem (GCP) is a well known $NP$-Hard problem which has received many attention from the scientific community because of its large range of real applications and computational difficulty. 
Given a simple graph $G = (V, E)$ a $k$-\emph{coloring} is a partition of $V$ into $k$ non-empty stable sets called \emph{color classes}, denoted by $C_1, \ldots, C_k$, such that vertices in $C_i$ are colored with color $i$ for $i \in  \{ 1,\ldots,k \}$.
GCP consists of finding the \emph{chromatic number} of $G$, denoted by $\chi(G)$, which is the minimum number $k$ of colors such that $G$ admits a $k$-coloring.

There are many practical situations that can be modeled as a GCP with some additional restrictions. For instance, in scheduling problems, it
would be desirable to assign a balanced workload among employees to avoid unfairness, or to assign balanced task load to prevent further
wear on some machines more than others.
The addition of this extra \emph{equity} constraint gives rise to the \emph{Equitable Coloring Problem} (ECP).

Formally, an \emph{equitable $k$-coloring} (or just $k$-eqcol) of $G$ is a $k$-coloring satisfying the \emph{equity constraint},
\ie $\left| |C_i| - |C_j| \right| \leq 1$, for $i, j \in \{1, \ldots, k \}$ or, equivalently,
$\lfloor n / k \rfloor \leq |C_j| \leq \lceil n / k \rceil$ for each $j \in \{1, \ldots, k \}$, where $n = |V|$.
The \emph{equitable chromatic number} of $G$, $\chi_{eq}(G)$, is the
minimum $k$ for which $G$ admits a $k$-eqcol. The ECP consists of finding $\chi_{eq}(G)$ and it is also an $NP$-hard problem \cite{KUBALE}.

One of the most well known exact algorithms for GCP is Branch-and-Bound \textsc{DSatur}, proposed by Br\'elaz in \cite{DSATUR}
and later improved by Sewell in \cite{SEWELL}. This algorithm is still used by its simplicity, its efficiency in medium-sized graphs and the possibility of applying it at some stage in metaheuristics or in more
complex exact algorithms like Branch-and-Cut. Recently, it was shown that a modification of \textsc{DSatur} performs relatively well compared with many
state-of-the-art Branch-and-Cut algorithms, showing superiority in random instances \cite{PASS}. This fact makes nowadays research on DSatur-based solvers
still important.

The goal of this work is to present a DSatur-based solver for the ECP. In the next sections, we 
review \textsc{DSatur} algorithm and we propose new pruning rules specifically derived from the equity constraint. Finally, we report our computational experience.

\vspace{-10pt}
\subsection{Notations and the \textsc{DSatur} algorithm}

\textsc{DSatur} is an implicit enumeration algorithm 
where each node of the tree corresponds to a
partial coloration of the graph. 

A \emph{partial} $k$-\emph{coloring} of $G$, $\Pi = (k, C_1, \ldots, C_n, U, F)$, is defined by a positive integer $k$, a family of disjoint stable sets  $C_1, \ldots, C_n$ of $G$ such that $C_j \neq \varnothing$ if and only if $j \leq k$,  \emph{a set of uncolored vertices} $U = V \backslash (\cup_{j=1}^k C_j)$ and
a list $F$ of their \emph{feasible color sets}, \ie
for every $u\in U$, $F(u)=\{j\in\{1,\ldots,n\} : \textrm{no vertex of}~C_j~\textrm{is adjacent to}~u \}$.
Clearly, a partial $k$-coloring with $U=\varnothing$ is a $k$-coloring.

Given $\Pi = (k, C_1, \ldots, C_n, U, F)$, $u \in U$ and $j \in \{1,\dots,k+1\}$, we denote by $\langle u,j \rangle \hookrightarrow \Pi$ to the partial coloring of $G$ obtained by adding node $u$ to $C_j$, \ie if
$\langle u,j \rangle \hookrightarrow \Pi = (k', C'_1, \ldots, C'_n, U', F')$ then $k' = \max \{j,k\}$, $C'_j = C_j \cup \{u\}$,
$C'_r = C_r$ for all $r \neq j$, $U' = U \backslash \{u\}$, $F'(v) = F(v) \cap \{j\}$ for each $v \in U'$ adjacent to $u$,
and $F'(v) = F(v)$ otherwise.

Given a maximal clique $Q = \{v_1, v_2$, $\ldots$, $v_q\}$ of $G$, we denote by $\Pi_Q$ the partial $q$-coloring defined by 
$C_i = \{v_i\}$ for all $1 \leq i \leq q$ and $U = V \backslash Q$. 

\textsc{DSatur} is based on a generic enumerative scheme proposed by Brown \cite{BROWN}, outlined as follows:

\medskip

{ \footnotesize
\noindent \underline{\textsc{Initialization}}: $G$ a graph, $\overline{c}$ an $UB$-coloring of $G$ and $Q$ a maximal clique of $G$.\\
\noindent \underline{\textsc{Node}$(\Pi = (k, C_1, \ldots, C_n, U, F))$}:~~~~~($UB$ and $\overline{c}$ are global variables)\\
\indent \emph{Step 1}. If $U = \varnothing$, set $UB \leftarrow k$, $\overline{c} \leftarrow \Pi$ and return.\\
\indent \emph{Step 2}. Select a vertex $u \in U$.\\
\indent \emph{Step 3}. For each color $1 \leq j \leq \max\{k+1,UB-1\}$ such that $j \in F(u)$, do: \\
\indent \indent $\Pi' \leftarrow \bigl( \langle u,j \rangle \hookrightarrow \Pi \bigr)$\\
\indent \indent If $F'(v) \neq \varnothing$ for all $v \in U'$, execute \textsc{Node}$(\Pi')$.
}

\medskip

It is not hard to see that the recursive execution of \textsc{Node}$(\Pi_Q)$ finally gives the value of $\chi(G)$ into the variable $UB$ and an optimal coloring into $\overline{c}$ \cite{DSATUR}.

Based on this scheme, different algorithms for solving GCP have been evaluated by proposing different \emph{vertex selection strategies} in Step 2.
Two of them are due to Br\'elaz (\textsc{DSatur} algorithm \cite{DSATUR}) and Sewell (\textsc{Celim} algorithm \cite{SEWELL}).
Recently, San Segundo proposed an improvement to Sewell rule (\textsc{Pass} algorithm \cite{PASS}), which gave rise to a competitive solver
with respect to many of the recent exact algorithms in the literature.

\vspace{-10pt}
\section{\textsc{EqDSatur}: An exact algorithm for the ECP} \label{SEQDSATUR}

\vspace{-5pt}
Let us notice that a trivial DSatur-based algorithm for the ECP can be obtained from the previous Brown's scheme, by 
changing the $UB$-coloring in the initialization by an $UB$-eqcol and adding the condition ``$\Pi$ is an equitable coloring'' in Step 1.
However, using explicitly the equity property during generation of nodes we can avoid to explore tree regions that will not lead to an equitable coloring and therefore would be needlessly enumerated.

In the following lemma, we present necessary conditions for a partial coloring to be extended to an equitable coloring. 

\begin{lemma}
Let $G$ be a graph of $n$ vertices and let $UB$ and $LB$ be, respectively, an upper and a lower bound of $\chi_{eq}(G)$. Let $\Pi$ be
a partial $k$-coloring of $G$ such that $k < UB$ 
and $M = max\{|C_r|: 1 \leq r\leq k\}$.
Then, if $\Pi$ can be extended to an $r$-eqcol of $G$ with $k \leq r  < UB$, the following properties hold:\\
\noindent (P.1) $|U| \geq \sum_{r=1}^k (\max\{M-1, \left\lfloor \frac{n}{UB-1}\right\rfloor\} - |C_r|)^+$
~~~~~~~ (P.2) $M \leq \left\lceil \frac{n}{\max\{k,LB\}} \right\rceil$
\end{lemma}

\vspace{-5pt}
In addition, property P.1 in the previous lemma also gives us a sufficient condition for a partial coloring with $U=\varnothing$ to be an equitable coloring:

\vspace{-5pt}
\begin{lemma}
If $\Pi$ is a partial $k$-coloring satisfying property P.1 and $U=\varnothing$ then $\Pi$ is a $k$-eqcol.
\end{lemma}
\vspace{-5pt}
Our DSatur-based algorithm introduces the previous properties as modifications (written in boldface) into the Brown's scheme, in the following way:

\medskip

{ \footnotesize
\noindent \underline{\textsc{Initialization}}: $G$ a graph, $\overline{c}$ an $UB$-{\bf eqcol} of $G$, {\bf LB a lower bound of $\chi_{eq}(G)$}
and $Q$ a maximal clique of $G$.\\
\noindent \underline{\textsc{Node}$(\Pi = (k, C_1, \ldots, C_n, U, F))$}:~~~~~($UB$ and $\overline{c}$ are global variables)\\
\indent \emph{Step 1}. If $U = \varnothing$, set $UB \leftarrow k$, $\overline{c} \leftarrow \Pi$ and return.\\
\indent \emph{Step 2}. Select a vertex $u \in U$.\\
\indent \emph{Step 3}. For each color $1 \leq j \leq \max\{k+1,UB-1\}$ such that $j \in F(u)$, do: \\
\indent \indent $\Pi' \leftarrow \bigl( \langle u,j \rangle \hookrightarrow \Pi \bigr)$\\
\indent \indent If $F'(v) \neq \varnothing$ for all $v \in U'$ {\bf and $\Pi'$ satisfies P.1 and P.2}, execute \textsc{Node}$(\Pi')$.
}

\medskip

We have the following:

\vspace{-5pt}
\begin{theorem}
The recursive execution of \textsc{Node}$(\Pi_Q)$ gives the value of $\chi_{eq}(G)$ into the variable $UB$
and an optimal equitable coloring into $\overline{c}$. 
\end{theorem}

\vspace{-5pt}
Proofs of the previous results are omitted due to lack of space.

In order to initialize our algorithm we use the heuristic \textsc{Naive} given in \cite{KUBALE} for obtaining an initial $UB$-eqcol of $G$. The lower bound $LB$ is computed as in \cite{DANIEL}, \ie as the maximum between the size of the maximal clique computed greedily and a relaxation of a bound given
in \cite{EQTREE}.

Regarding the mentioned vertex selection strategies, we carried out benchmark tests over random instances  
and we concluded 
that \textsc{Pass} \cite{PASS} is the best choice. We call $\textsc{EqDSatur}$ to our implementation using this strategy.

Another factor we take into account is 
in which order the nodes are evaluated in Step 3. 
All mentioned DSatur implementations for GCP evaluate  
first color $j=1$, then color $j=2$, and so on. We call $\textsc{EqDS}_1$ to
$\textsc{EqDSatur}$ with this criterion. In addition, we consider another strategy based on sorting color classes 
according to their size in ascending order: if $|C_{i_1}| \leq |C_{i_2}| \leq \ldots \leq |C_{i_k}|$, we evaluate first $j=i_1$, then $j=i_2$, and so on.
We call $\textsc{EqDS}_2$ to $\textsc{EqDSatur}$ with this strategy.

\vspace{-10pt}
\section{Computational experience}

\vspace{-5pt}
Computational tests were carried out on an Intel i5 2.67Ghz over Linux O.S.
Some details and tables were omitted due to lack of space. Instead, a summary of the most essential details are provided.

Our first experiment compares $\textsc{EqDS}_1$ against the ``trivial'' DSatur-based exact algorithm for the ECP mentioned at Section \ref{SEQDSATUR}, in order to evaluate 
whether 
the time needed to check properties P.1 and P.2 compensate for 
the time wasted in
exploring nodes of the enumeration tree where these properties do not hold. We noticed that $\textsc{EqDS}_1$ really outperforms the trivial implementation.
For instance, in medium-density random graphs of 70 vertices, $\textsc{EqDS}_1$ is in average 25 times faster than the trivial implementation and
is able to solve 21\% more instances within 2 hours of execution.

The second experiment compares $\textsc{EqDS}_1$ and $\textsc{EqDS}_2$ against ``integer linear programming-based'' solvers for ECP, a classical
approach for developing exact algorithms.
We consider the recent Branch-and-Cut proposed in \cite{BYCBRA} (B\&C-$LF_2$) for which the authors report results on random instances up to 70 vertices in a 1.8 Ghz AMD-Athlon platform.
For random instances with 80 vertices, we compare \textsc{EqDSatur} against CPLEX 12.1 solving the formulation for ECP given in \cite{DANIEL}
with the same initial bounds. 

The following table 
reports the percentage of solved instances, average of relative gap and time elapsed for 100 random instances
(10 per row, except results from \cite{BYCBRA}). An instance is considered \emph{not solved} after the limit of two hours of execution.
A bar ``$-$'' means no instance was solved. Each instance of $d \%$ of density is generated by considering a uniform probability $d$ that two
vertices are adjacent to each other.

\begin{center} 
\tiny
\begin{tabular}{c@{\hspace{4pt}}c|c@{\hspace{4pt}}c@{\hspace{4pt}}c|c@{\hspace{4pt}}c@{\hspace{4pt}}c|c@{\hspace{4pt}}c@{\hspace{4pt}}c}
  & \%Density & \multicolumn{3}{c|}{\% solved inst.} & \multicolumn{3}{c|}{\% Rel. Gap.} & \multicolumn{3}{c}{Time} \\
Vertices & Graph & $LF_2$ & $\textsc{EqDS}_1$ & $\textsc{EqDS}_2$ & $LF_2$ & $\textsc{EqDS}_1$ & $\textsc{EqDS}_2$ & $LF_2$ & $\textsc{EqDS}_1$ & $\textsc{EqDS}_2$ \\
\hline
70 & 10 & 100 & 100 & 100 & 0   & 0 & 0 & 109 & 0  & 0 \\
70 & 30 &  0  & 100 & 100 & 18  & 0 & 0 & $-$ & 0  & 0 \\
70 & 50 &  0  & 100 & 100 & 8,2 & 0 & 0 & $-$ & 16,2 & 16,1 \\
70 & 70 & 100 & 100 & 100 & 0   & 0 & 0 & 273 & 31,2 & 32,1 \\
70 & 90 & 100 & 100 & 100 & 0   & 0 & 0 & 11  & 0  & 0 \\
\hline
 & & CPLEX & $\textsc{EqDS}_1$ & $\textsc{EqDS}_2$ & CPLEX & $\textsc{EqDS}_1$ & $\textsc{EqDS}_2$ & CPLEX & $\textsc{EqDS}_1$ & $\textsc{EqDS}_2$ \\
\hline
80 & 10 & 100 & 100 & 100 &  0 &  0 &  0 & 5,7  & 0    & 0 \\
80 & 30 &  0  & 100 & 100 & 20 &  0 &  0 & $-$  & 23,7 & 17,6 \\
80 & 50 &  0  & 100 & 100 & 24 &  0 &  0 & $-$  & 477  & 424 \\
80 & 70 &  10 &  70 &  70 & 16 & 10 & 10 & 5769 & 1715 & 1653 \\
80 & 90 & 100 & 100 & 100 &  0 &  0 &  0 & 690  & 12,5 & 12,3 \\
\hline
\end{tabular}
\end{center}
\smallskip

Our algorithm is able to solve more instances than CPLEX and, without considering the difference of platforms, B\&C-$LF_2$.
Also, $\textsc{EqDS}_2$ seems to be a little better than $\textsc{EqDS}_1$ in terms of time.

The last experiment compares $\textsc{EqDS}_1$ and $\textsc{EqDS}_2$ against CPLEX and B\&C-$LF_2$ on
24 benchmark instances reported in \cite{BYCBRA}.
16 instances have been solved by CPLEX, $\textsc{EqDS}_1$ and $\textsc{EqDS}_2$ in less than two seconds.
B\&C-$LF_2$ solved these 16 instances with an average of 62 seconds and never outperforms the other three algorithms.
Instance \texttt{queen8\_8} have been solved by CPLEX in 654 sec., by B\&C-$LF_2$ in 441 sec., by $\textsc{EqDS}_1$ in 7.5 sec. and by $\textsc{EqDS}_2$ in 1.1 sec. Instances \texttt{miles1000} and \texttt{miles750} have not been solved by
$\textsc{EqDS}_1$, but have been solved by B\&C-$LF_2$ in 267 and 171 sec. respectively, and by CPLEX and $\textsc{EqDS}_2$ in less than a second.

On the other hand, neither $\textsc{EqDS}_1$ nor $\textsc{EqDS}_2$ could solve 5 instances.
In particular, 3 of these instances (\texttt{3-FullIns\_3}, \texttt{4-FullIns\_3}, \texttt{5-FullIns\_3}) are hard to solve by enumerative
schemes, as reported in \cite{PASS}, so in our opinion, \textsc{EqDSatur} presents the expected behaviour.

We want to remark that \textsc{EqDSatur} has also been able to solve DIMACS benchmark instances which are not mentioned in \cite{BYCBRA}.
For instance, \texttt{queen9\_9} has been solved by $\textsc{EqDS}_1$ and $\textsc{EqDS}_2$ in less than 10 minutes. In contrast, CPLEX
could not solve it in the term of 4 hours of execution. Another example is \texttt{myciel5} which
has been solved in less than a sec.\! by $\textsc{EqDS}_1$ and $\textsc{EqDS}_2$ but CPLEX needed 149 sec.\! to solve it. 

From these results, we conclude that $\textsc{EqDS}_2$ is highly competitive with respect to the algorithms available in the literature for the ECP.


\end{document}